\documentclass[%
 reprint,
superscriptaddress,
 amsmath,amssymb,
aps,
prapp,
longbibliography,
]{revtex4-1}

\usepackage[utf8]{inputenc}
\usepackage[english]{babel}
\usepackage[T1]{fontenc}
\usepackage{hyperref}
\usepackage{amsmath}
\usepackage{amssymb}
\hypersetup{colorlinks=true,
    linkcolor=blue,
    filecolor=blue, 
    citecolor=blue,
    urlcolor=blue,
}

\usepackage{braket, hyperref, mathtools, verbatim}
\usepackage{physics}
\usepackage{graphicx}
\usepackage{braket}
\usepackage{dcolumn}
\usepackage{bm}
\usepackage{tabularx}
\usepackage{makecell}
\usepackage{tikz}
\usepackage{lipsum}
\usepackage{bbold}
\usepackage{enumerate}

\newcommand{\Z}{\mathbb{Z}}

\newcommand{\sE}{\mathcal{E}}

\newcommand{\sL}{\mathcal{L}}
\newcommand{\sH}{\mathcal{H}}

\newcommand{\dphi}{\dot{\phi}}
\newcommand{\ddq}{\dot{q}}
\newcommand{\ddp}{\dot{p}}

\allowdisplaybreaks

\begin{document}

\title{On-demand single-microwave-photon source in a superconducting circuit with wideband frequency tunability}

\author{Samarth Hawaldar}
\email{samarth.hawaldar@ist.ac.at}
\affiliation{Department of Instrumentation and Applied Physics, Indian Institute of Science, Bengaluru - 560012, India}
\affiliation{Institute of Science and Technology Austria, 3400 Klosterneuburg, Austria}
\author{Siddhi Satish Khaire}
\affiliation{Department of Instrumentation and Applied Physics, Indian Institute of Science, Bengaluru - 560012, India}
\author{Per Delsing}
\affiliation{Department of Microtechnology and Nanoscience, Chalmers University of Technology, 41296 Gothenburg, Sweden}
\author{Baladitya Suri}
\email{surib@iisc.ac.in}
\affiliation{Department of Instrumentation and Applied Physics, Indian Institute of Science, Bengaluru - 560012, India}

\begin{abstract}
In this article, we propose a new method of generating single microwave photons in  superconducting circuits. We theoretically show that pure single microwave photons can be generated on demand and tuned over a large frequency band by making use of Landau-Zener transitions under a rapid sweep of a control parameter. We devise a protocol that enables fast control of the  frequency of the emitted photon over two octaves, without requiring extensive calibration. Additionally, we make theoretical estimates of the generation efficiency, tunability, purity, and linewidth of the photons emitted using this method for both charge and flux qubit-based architectures. We also provide estimates of optimal device parameters for these architectures in order to realize the device.
\end{abstract}

\maketitle

\section{\label{sec:Introduction}Introduction}
The field of Quantum Information Science (QIS) deals with the encoding and manipulation of information in quantum objects\cite{eisamanInvitedReviewArticle2011}. Communication of quantum information between spatially separated computing/storage nodes, i.e. between qubits, is an important component of QIS \cite{pechalMicrowaveControlledGenerationShaped2014}. Photonic qubits are an ideal choice for this role \cite{eisamanInvitedReviewArticle2011,forn-diazOnDemandMicrowaveGenerator2017} because of their transmission at the speed of light, weak interaction with the environment leading to robustness against external noise, and the ease of using linear optics for spatial manipulation. In this context, single-photon sources find use in Quantum computation \cite{knillSchemeEfficientQuantum2001, zhongQuantumComputationalAdvantage2020, kokLinearOpticalQuantum2007}, communication \cite{kimbleQuantumInternet2008}, cryptography (primarily Quantum Key Distribution) \cite{inamoriUnconditionalSecurityPractical2007, bennettQuantumCryptographyPublic2014, ekertQuantumCryptographyBased1991,duanLongdistanceQuantumCommunication2001}, and Quantum sensing\cite{degenQuantumSensing2017}.\\

A practical single-photon source must have a high efficiency of generating photons on demand with frequencies tunable over a wide range \cite{pengTuneableOndemandSinglephoton2016}, and with a low timing {jitter} in their emission \cite{forn-diazOnDemandMicrowaveGenerator2017}. Additionally, it may have the ability to generate shaped photons \cite{pechalMicrowaveControlledGenerationShaped2014}, and should be compact so that multiple generators can be accommodated on a single chip\cite{LiPhyRevRes2024}. The most commonly employed method of generating single photons in any region of the electromagnetic spectrum is to excite a two-level system (TLS), natural or artificial, and then extract a photon \textit{via} spontaneous emission. In the optical regime, artificial two-level systems are realized using NV centres in diamond \cite{kurtsiefer_stable_2000}, single molecules \cite{brunel_triggered_1999}, and quantum dots \cite{hours_single_2003, WeiNanoLett2014} among many other candidates. In comparison, in order to realize a two-level system in the microwave regime, we require very low temperatures ($<100\,$mK) inside a dilution cryostat to avoid a thermal background of photons and to mitigate the effect of thermal excitations.  

The first successful realization of a single-photon source in the microwave regime \cite{houckGeneratingSingleMicrowave2007} made use of a transmon qubit \cite{koch_charge-insensitive_2007} coupled to a waveguide resonator. In \cite{houckGeneratingSingleMicrowave2007}, and in the majority of work that has followed it, the transmon was excited using a drive at frequency $\omega_{ge}$, the transition frequency from the ground to the first excited state. As a consequence, a single photon is spontaneously emitted at the same frequency. Houck \textit{ et al.} \cite{houckGeneratingSingleMicrowave2007} achieved a usable efficiency of 78\%, and fulfiled the requirements of being on-demand and, having low timing {jitter} due to its short excited-state lifetime ($T_1\approx90\,$ns). Pechal \textit{ et al.} \cite{pechalMicrowaveControlledGenerationShaped2014}, using a similar architecture to that in Ref. \cite{houckGeneratingSingleMicrowave2007}, generated shaped photons with a usable efficiency of $\approx 76\%$ by making use of the 2$^\text{nd}$ excited state of the transmon to generate a single excitation in the resonator. A further improvement to the emission band-width over \cite{houckGeneratingSingleMicrowave2007} was achieved by Peng \textit{et al.} \cite{pengTuneableOndemandSinglephoton2016} by using a flux qubit directly coupled to two transmission lines,  one for input and the other for output, without a resonator in the architecture. Peng \textit{et al.} reported a usable efficiency of $>65\%$ over a $3\,$GHz range. Improvement to the generation efficiency of photons was made by Zhou \textit{et al.} in \cite{zhouTunableMicrowaveSinglePhoton2020}, where the source geometry was engineered to emit photons with a peak efficiency of 87\% and a tunability of greater than 1GHz. In the aforementioned protocols (except by Pechal \textit{et al.} \cite{pechalMicrowaveControlledGenerationShaped2014}), the excitation of the qubit is caused by drive photons at the same frequency as the single photon at the output. Therefore, the output contains drive photons alongside the desired single photons, thereby reducing the purity of the single photon state at the output. Lu \textit{et al.} \cite{luQuantumEfficiencyPurity2021} mitigate this leakage by introducing a cancelling pulse at the input frequency to enhance the single-photon purity of the output. They reported a suppressed leakage of less than 0.005 photons, while maintaining a usable efficiency between 71-99\%. An alternate method of exciting the system makes use of stimulated Raman adiabatic passage (STIRAP) \cite{mangano_single_2008, premaratne_microwave_2017, yan_fast_2021}. This method also allows for fast preparation of a general superposition of Fock states, while allowing mitigation of environmental decoherence effects and leakage by driving at a frequency separate to the emitted photons'. In all of these methods, even in the ones where it is possible to tune the frequency of the emitted photon, there is the issue of needing to calibrate the drives in the system after changing the frequency of the emitter and that of the fact that the frequency tuning is a slow process in comparison to the photon generation time.

In this work, we propose a new protocol for generating high-purity single microwave photons on-demand with high efficiency over a large range of frequencies using a device with a very small footprint on the chip. Hence, this protocol fulfils all the criteria of good single-photon generators, except that of the ability to shape the emitted photons. In this protocol, we use diabatic Landau-Zener transitions instead of a coherent drive tone to excite two-level systems. Within the superconducting qubit platform, we consider charge qubits and flux qubits as candidate tunable sources of single photons using this protocol. We excite either of these systems using diabatic transitions caused by a rapid sweep of a control parameter across an avoided level-crossing. Since this protocol does not employ any coherent drive field, it circumvents the issue of qubit drive tone leaking into output and affecting the purity of the emitted single photon state. By maximizing the diabatic transition probability while keeping the relaxation time of the qubit short, an on-demand source with an emission rate in megahertz can be achieved. Additionally, we propose a new protocol that allows for a fast control of the frequency of the emitted photons \textit{in situ} over two octaves,  without needing to recalibrate the system in principle.  \\

This paper is structured as follows. In Sec.~\ref{sec:LZ}, we provide a review of Landau-Zener transitions and look at their applicability to Cooper-pair boxes and flux qubits. Then in Sec.~\ref{sec:Pulse}, we analyse a few pulse shapes and provide a pulse sequence for optimal device operation. Finally, in Sec.~\ref{sec:Est}, we provide estimates of device performance for realistic parameters and present a few parameter regimes for device implementation.

\section{\label{sec:LZ}Landau-Zener Transitions for Single Photon Generation}
\subsection{Landau-Zener Transitions}
\begin{figure}[!tb]
\includegraphics[width=0.9\columnwidth]{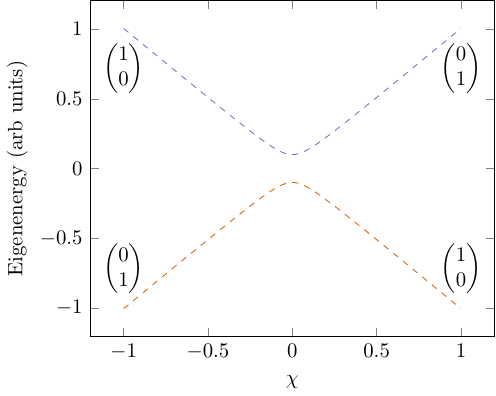}
\caption{Eigenvalues of the two-level LZ Hamiltonian, Eq.~\ref{eq:LZ_Ham_TLS}, as a function of the swept parameter $\chi=\hbar\beta t$ for $\hbar\Delta=0.1$ The eigenstates of the Hamiltonian have been labelled at the extrema}
\label{fig:LZ_Ham}
\end{figure}

The theory of Landau-Zener-Majorana-Stuckelburg transitions, often simply termed Landau-Zener (LZ) transitions, describes the transition probabilities of a system in which the eigenstates (and energy eigenvalues) are tuned using an external parameter. The theory predicts that  a rapid sweep of a parameter across an avoided crossing of two eigenstates (see Fig.~\ref{fig:LZ_Ham}) leads to a diabatic transition from one to the other. For a two-level system initially in the ground state, we start by considering a Hamiltonian with a linear time-dependence. An exact solution for this Hamiltonian was shown independently by Landau \textit{et al.} \cite{landauZurTheorieEnergieubertragung1932}, Zener  \cite{zenerNonadiabaticCrossingEnergy1932}, Stuckelburg \cite{stueckelbergTheorieUnelastischenStosse1932}, and Majorana \cite{majoranaAtomiOrientatiCampo1932} in 1932. In general, a Hamiltonian of this form for a TLS can be written as
\begin{equation}\label{eq:LZ_Ham_TLS}
    H = \hbar\mqty(\omega_0 & D/2 \\ D/2 & \omega_0 - \beta t) \equiv \frac{\hbar\beta t}{2}\hat{\sigma}_z + \frac{\hbar D}{2}\hat{\sigma}_x,
\end{equation}
for a time-independent $\beta$ where $\hat{\sigma}_x, \hat{\sigma}_z$ are Pauli matrices. In Eqn.~\ref{eq:LZ_Ham_TLS}, the middle and the final parts are not equal, mathematically speaking, but are related by a gauge transformation that subtracts the same value from each diagonal term keeping the details of the dynamics of the states the same up to a global phase. Starting in the ground state $\ket{g}$ at $t=-\infty$, solving the Schr\"{o}dinger equation with the above Hamiltonian, the probability of transition to the excited state $\ket{e}$ at $t=\infty$ is given by (see Appendix \ref{secApp:LZderivation} for details)
\begin{equation}\label{eq:ProbLZ}
    P_{g \rightarrow e} = \exp(-\frac{\pi D^2}{2\beta}).
\end{equation}

This problem can be generalised to study the evolution of an $N$-level system \cite{shytovLandauZenerTransitionMultilevel2004,demkovStationaryNonstationaryProblems1968, demkovExactSolutionMultistate2001, sinitsynSolvableMultistateModel2016} in the presence of a linearly time-dependent tuning of the Hamiltonian of the form
\begin{equation}\label{eq:LZ_Ham}
	H(t) = A + Bt,
\end{equation} 
where $A,B$ are $N\times N$ time-independent matrices. Given the Hamiltonian described in Eq.~\ref{eq:LZ_Ham}, and the fact that the system is in some eigenstate $\ket{i}$ at $t=-\infty$, the generalized LZ problem aims to compute the probability of the system being in some eigenstate $\ket{j}$ at $t=\infty$. For a general $N$-level system, an exact solution is not yet known, but for very specific cases solutions have been found\cite{demkovStationaryNonstationaryProblems1968, demkovExactSolutionMultistate2001, sinitsynSolvableMultistateModel2016}.

\begin{figure*}[!tb]
\centering
\includegraphics[width=\linewidth]{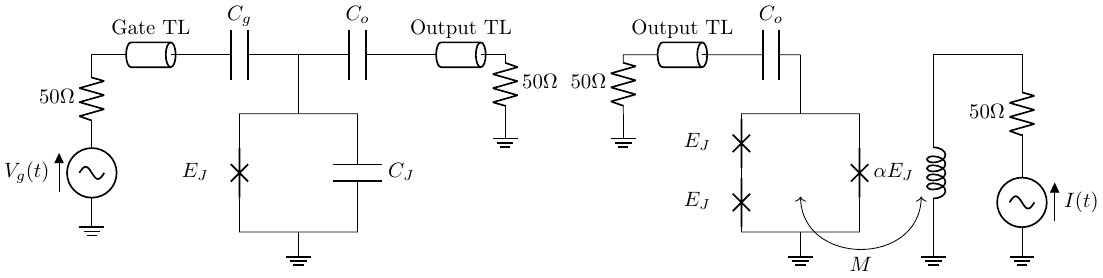}
 \caption{Schematic for CPB-based generator (left) and Flux qubit-based generator (right)}
 \label{fig:CPB_Flux_design}
\end{figure*}

\subsection{Cooper-pair box}
The Hamiltonian in Eq.~\ref{eq:LZ_Ham} can, for instance, be implemented in superconducting qubit architectures such as a Cooper-pair box (CPB) or a flux qubit. A CPB is a single Josephson junction (JJ) with an associated intrinsic capacitance ($C_J$), connected to an external bias-voltage source $V_g$   \textit{via} a capacitance $C_g$ \cite{shnirman_quantum_1997, makhlin_quantum-state_2001} (see Fig.~\ref{fig:CPB_Flux_design}).

In the quantized charge basis $\ket{n}$ where $n$ is the difference of the number of Cooper-pairs between the two CPB electrodes, the Hamiltonian of the CPB can be written as \cite{shnirman_quantum_1997, makhlin_quantum-state_2001, bladh_single_2005},
\begin{multline}\label{eq:HamCPB}
    \hat{H} = \sum_{n\in\Z} \Big(E_Q (n-n_g)^2 \op{n}{n} \\\left.- \frac{E_J}{2} \left( \op{n}{n+1} + \op{n+1}{n} \right)\right),
\end{multline}
where $E_Q=2e^2/C_\Sigma$ is the electrostatic energy of an effective capacitance $C_\Sigma$ with one Cooper-pair of charge on it, $E_J$ is the Josephson energy associated with the tunnelling of Cooper-pairs across the junction, and $n_g=(C_g V_g)/(2e)$ is the normalised gate charge.

Dropping the terms independent of $n$ \footnote{This is allowed because the form of the $n$-independent terms is $n_g^2 \sum_n \op{n}{n} = n_g^2 \cdot \mathbb{1}$. This term can be thought of as just a global reference potential offset which does not affect either the eigenstates or the transition frequencies of the Hamiltonian being studied.}, the Hamiltonian can be rewritten in the LZ form by considering $n_g = \lambda t$  as the swept parameter, where $\lambda = C_g \dot{V_g}/ (2e)$ for a constant $\dot{V_g}$. This gives
\begin{multline}\label{eq:HamCPB_LZ}
    \hat{H} = \sum_{n\in\Z} \left(E_Q n^2\op{n}{n} - \frac{E_J}{2} \left( \op{n}{n+1} + \op{n+1}{n} \right)\right)\\*
    - 2E_Q \lambda t \sum_{n\in\Z} n \op{n}{n} = A + Bt.
\end{multline}
When $E_Q \gg E_J$, by constraining $n_g \in [0,1]$ \footnote{This is equivalent to choosing any other range of $n_g \in [m+\epsilon_g,m+1-\epsilon_g]$ for $m\in\Z$ for which the subspace of interest would be $n=m,m+1$}, the system can be approximated to a TLS with a Hamiltonian given by
\begin{equation}\label{eq:HamCPB_LZ_TLS}
    \hat{H} = \mqty(0 & -E_J/2 \\ -E_J/2 & E_Q - 2E_Q\lambda t) \equiv E_Q \lambda t \hat{\sigma}_z - \frac{E_J}{2} \hat{\sigma}_x,
\end{equation}
where the $n_g$ has been redefined to be given by $n_g = \lambda t + 0.5$ such that the avoided crossing occurs at $t=0, n_g=0.5$. We note that, more practically, as the probability of leakage due to LZ transition depends strongly on how close to the avoided crossing the system is, by constraining $n_g \in [\epsilon_g, 1-\epsilon_g]$ for large enough $\epsilon_g$ such that the leakage outside the $n=0,1$ subspace can be neglected (estimates on how large $\epsilon_g$ must be can be obtained from Fig.~\ref{fig:leakage_vs_pulserange} in Appendix \ref{secApp:Leakage}), the TLS approximation can be achieved.
Comparing Eq.\ref{eq:HamCPB_LZ_TLS} to Eq.~\ref{eq:LZ_Ham_TLS}, we have for this system $\hbar\beta = 2 E_Q \lambda$. Hence, for this device, making use of Eq.~\ref{eq:ProbLZ}, we get the probability of excitation to be,
\begin{equation}\label{eq:ProbCPB}
    P_\text{ex,CPB} = \exp(-\frac{\pi E_J^2}{4\hbar E_Q \lambda}).
\end{equation}

We can also calculate the transition frequency as a function of $n_g$ to be given as
\begin{equation}
    \hbar \omega_\text{CPB} = \sqrt{E_J^2 + E_Q^2 (1 - 2n_g)^2}.
\end{equation}

\subsection{Flux Qubits}
Another superconducting architecture in which LZ transitions can be implemented is the flux qubit. A standard flux qubit consists of three junctions in a single loop, two of which are identical with critical current $I_c$ (and Josephson energy $E_J$), and the third one has critical current $I_c/\alpha$ (and Josephson energy $\alpha E_J$) where $\alpha$ is the junction asymmetry (see Fig.~\ref{fig:CPB_Flux_design}). The two identical junctions have an intrinsic capacitance $C$ each, and the third junction is shunted with a capacitance $C_s$. The Hamiltonian for this system close to the optimal point where the external flux $\phi_e$ can be written as $\phi_e = (n+0.5)\Phi_0 + \delta\phi_e$ ($\Phi_0 = h/2e$ is the magnetic flux quantum, and half-integer multiples of $\Phi_0$ are avoided crossings in the flux-qubit spectrum), can be expressed approximately in terms of a tunneling energy $\Delta$ \cite{mooij_josephson_1999, robertson_quantum_2006, deppe_superconducting_2009} as,
\begin{equation}\label{eq:FluxHamApprox}
    \hat{H} = -\gamma I_c \delta\phi_e \hat{\sigma}_z - \frac{\Delta}{2} \hat{\sigma}_x,
\end{equation}
where $\gamma = \sqrt{1 - \left( \frac{1}{2\alpha} \right)^2}$.

For $\delta\phi_e = \frac{\Phi_0}{2\pi}\mu t$, we have once again a LZ type Hamiltonian,
\begin{equation}\label{eq:FluxHamLZ}
    \hat{H} = - \frac{\Delta}{2} \hat{\sigma}_x -\gamma E_J\mu t \hat{\sigma}_z,
\end{equation}
with transition frequency given by
\begin{equation}\label{eq:FluxTrans}
    \hbar \omega_\text{flux} = \sqrt{(2\gamma I_c \delta\phi_e)^2 + \Delta^2}.
\end{equation}
Comparing to Eq.~\ref{eq:LZ_Ham_TLS}, we once again have for this system $\hbar\beta = 2\gamma E_J \mu$
For this system, the probability of LZ excitation is
\begin{equation}\label{eq:ProbFlux}
    P_\text{ex,flux} = \exp(-\frac{\pi\Delta^2}{2\hbar \gamma E_J \mu}).
\end{equation}

\section{\label{sec:Pulse} Protocol for Diabatic Excitations}
As can be seen from the expressions Eq.~\ref{eq:ProbLZ}, Eq.~\ref{eq:ProbCPB}, and Eq.~\ref{eq:ProbFlux}, by rapidly sweeping the control parameter across an avoided crossing, a qubit can be excited (with a high probability) due to a diabatic transition. We propose to use this phenomenon to excite a superconducting qubit (charge or flux) without an input microwave excitation. The probability of excitation is closest to unity when the control parameter is swept at the greatest possible rate within the limits set by experiment. In particular, from Eq.~\ref{eq:ProbLZ}, we see that this condition is satisfied  when $\beta \gg D^2$, which translates to the conditions  $\lambda \gg E_J^2/\hbar E_Q$ for the charge qubit and $\mu \gg \Delta^2/\hbar\gamma E_J$ for the flux qubit. 

At the outset, if a function generator or arbitrary waveform generator (AWG) capable of providing a fixed large sweep rate (slew-rate) with control over the initial and final points were available, the optimal pulsing sequence would just involve a sweep of the control parameter $\chi$ (for a CPB, $\chi$ is $(n_g-0.5)$, while for a flux qubit, $\chi$ is $\delta\phi_e$) from the initial value $\chi_i$ to the final (target) value $\chi_f$ at the maximum possible rate. However, most readily available AWGs  define a minimum rise time ($t_r$) for a fixed voltage range of sweep instead of a fixed slew rate. The minimum rise-time ($t_r$), along with the maximum output voltage of the AWG, defines the maximum slew-rate possible with that instrument. This also means that a sweep-range smaller than the maximum output range will result in a lower slew-rate. To achieve a high slew-rate, and thereby a high diabatic transition probability, we need to execute a sweep over the full range of interest of the control parameter $\chi$ in time $t_r$. One way to achieve this is to add a voltage divider between the AWG and the system such that a full-range sweep on the AWG corresponds to the sweep from $\chi_i$ to $\chi_f$ at the device. However, this method requires us to work with a fixed $\chi_i$ and $\chi_f$ for the whole experiment. Another way to achieve an optimal slew-rate, and thereby a maximal diabatic transition probability, is using the ``catapult" protocol we propose here.  Using this protocol, the maximum sweep rate can be achieved independent of the values of $\chi_i$ and $\chi_f$. This provides us pulse-level control on the values of $\chi_i$ and $\chi_f$ and, in turn, allowing for a pulse-level control on the frequency of the emitted photon.

\begin{figure}[h]
	\centering
	\includegraphics[width=0.9\columnwidth]{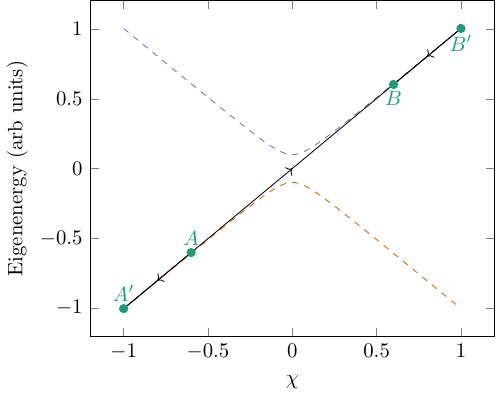}
	\caption{Pulse sequence for changing $2\chi/\chi_0$ from $-0.4$ to $0.7$ while performing a transition with the greatest probability afforded by the instrument}
	\label{fig:PulseSeq}
\end{figure}

According to this catapult protocol, we propose sweeping $\chi$ from its initial value $\chi_i$ ($A$ in Fig.~\ref{fig:PulseSeq}) to the minimum value $-\chi_0/2$ ($A'$ in Fig.~\ref{fig:PulseSeq}) in $t_r$, followed by a sweep from $A'$ to the maximum value $+\chi_0/2$ ($B'$ in Fig.~\ref{fig:PulseSeq}) in $t_r$, and finally to the target value $\chi_f$ ($B$ in Fig.~\ref{fig:PulseSeq}) in $t_r$. In effect, the control parameter is first pulled back to a minimum value, and then catapulted over the avoided crossing rapidly, in the minimum possible time $t_r$, to the maximum value.  We note that the transition happens only in the second step ( $A' - B'$ in Fig.\ref{fig:PulseSeq}) of the protocol, where we traverse across the avoided crossing with an associated rate of $\chi_0/t_r$.  Therefore, even though the total duration of the protocol is $3t_r$, the maximum slew-rate, and in turn the transition probability, are not affected. The single photon is then spontaneously emitted at the frequency determined by the final value $\chi_f$. Upon emission, the system returns to ground state with control parameter at $\chi_f$. The catapult protocol can now be implemented in reverse to return to the initial control parameter $\chi_i$, while emitting another photon, this time at frequency determined by $\chi_i$.   An important point to note here is that the protocol does not enforce any specific relation between the initial and final points, $\chi_i$ and $\chi_f$ respectively. This allows for fast control of the emitted photon frequency in each half-cycle of the catapult protocol.  

We now consider the question of the pulse-shape and its effects on the transition-probability. We consider four pulse-shapes, \textit{viz} linear, gaussian, hyperbolic tangent and exponential, and through numerical calculations we show that a linear sweep is sufficient for every step of the procedure ( see Appendix \ref{secApp:PulseShape} for details). 

It is important to note that the CPB and the flux qubit have a finite anharmonicity and therefore are not pure two-level systems. Therefore, we consider  the effect of the rapid sweep of the control parameter on the probability of excitation of the second and higher excited states of the multi-level system. Through numerical simulations we observe that, for a given $t_r$,  the leakage into the higher levels, which can lead to some multi-photon component in the emission, can be reduced by slightly  reducing the range of the sweep $A'-B'$ (see Fig.\ref{fig:PulseSeq}).  The resultant trade-off does not severely affect the transition probability to the first excited state ( see Appendix~\ref{secApp:Leakage} for details).

Finally, we also consider the effect of reactive circuit elements, namely $C_g$ for the charge qubit, or the coupling inductance $M$ for the flux qubit, on the actual pulse-shape that is transmitted to the device from the control transmission line. We note that, for usual experimentally feasible parameters, the time-constants associated with the reactive coupling elements ($Z_0 (C_g+C_o)$ for charge qubit and $M/Z_0$, where $Z_0 = 50\,\Omega$ is impedance of the environment) are in the femtosecond to picosecond range. Since the minimum rise-time of  AWGs currently available is in the hundreds of picosecond to nanosecond range, we conclude that this effect can be ignored. 

\subsection{Usable Efficiency}
For a single photon generator to be practically usable, the emitted photon needs to enter the output transmission line with a high probability. Hence, we define the usable efficiency $\eta$ as the probability that a photon is generated and emitted into the output line,
\begin{equation}\label{eq:UsableEfficiency}
    \eta = P(\text{Excitation})P(\text{Emission into Output})
        = P_\text{ex} \frac{\Gamma_o}{\Gamma_\text{tot}},
\end{equation}
where $\Gamma_o$ is the decay (emission) rate into the output line and $\Gamma_\text{tot} (= \Gamma_g + \Gamma_o + \Gamma_\text{nr})$ is the total decay rate of the TLS. For the below calculation, we assume that non-radiative decay is given by $\Gamma_\text{nr}$. 

For a CPB, using the effective coupling capacitances to the gate and output lines $C_{g,\text{eff}}, C_{o,\text{eff}}$ respectively (for the usual operating parameters, $C_{g,\text{eff}} \approx C_g, C_{o,\text{eff}}\approx C_o$. For the form of the expressions, see Eqn.~\ref{eqn:Ceff}), and performing the Master equation calculations for the complete Hamiltonian, we find the photon emission rate into the gate and output lines, $\Gamma_{g/o} \propto C_{g/o, \text{eff}}^2$ with equal proportionality constants. Hence, the usable efficiency $\eta_\text{CPB}$ is given by
\begin{equation}\label{eq:EffCPB}
    \eta_\text{CPB} = \frac{P_\text{ex,CPB} \Gamma_o}{\Gamma_g + \Gamma_o + \Gamma_\text{nr}} = \frac{\exp(-\frac{\pi E_J^2}{4\hbar E_Q \lambda})}{1 + (C_{g,\text{eff}}/C_{o,\text{eff}})^2 + \Gamma_\text{nr}/\Gamma_o}.
\end{equation}

For a flux qubit, we assume that the channels of decay are to the output line and to the flux line. Assuming the capacitive coupling to the output line $C_o$ and the mutual inductance with the flux line $M$, the decay rates are given by $\Gamma_o \propto \Delta C_o^2 E_J^2$ \cite{pengTuneableOndemandSinglephoton2016} and $\Gamma_f \propto M^2 E_J^2 \Delta^2 / \omega_\text{flux}$ \cite{deppe_superconducting_2009, koch_charge-insensitive_2007} respectively with different proportionality constants, say $K_o, K_f$. Hence, the usable efficiency $\eta_\text{flux}$ is
\begin{align}\label{eq:EffFlux}
    \eta_\text{flux} &= \frac{P_\text{ex,flux}}{1 + \Gamma_f/\Gamma_o + \Gamma_\text{nr}/\Gamma_0} \nonumber \\
    &= \frac{\exp(-\frac{\pi\Delta^2}{2\hbar E_J \mu \sqrt{1 - \left( \frac{1}{2\alpha} \right)^2}})}{1 + \frac{K_f M^2 \Delta}{K_o C_o^2 \omega_\text{flux}} + \frac{\Gamma_\text{nr}}{K_o \Delta C_o^2 E_J^2}}.
\end{align}

For usual experimental values of $M = 0.015\,\Phi_0/$mA \cite{koch_charge-insensitive_2007}, in the regime of high emission probability, $\Gamma_f \sim 10^3\,$s$^{-1}$ for $\Delta/h = 4\,$GHz (and $\Gamma_f \sim 10^2\,$s$^{-1}$ for $\Delta/h = 1\,$GHz) at $\omega_\text{flux} = 2\pi \times 4\,$GHz, which is much lower than the $\Gamma_o \sim 10^7\,$s$^{-1}$ which would be designed for a high repetition-rate device. Hence, we neglect this effect in further calculations, and for a flux qubit consider
\begin{equation}
	\eta_\text{flux} = P_\text{ex,flux}	\frac{\Gamma_o}{\Gamma_o + \Gamma_\text{nr}}.
\end{equation}

\section{\label{sec:Est}Estimates, Recommended Parameters, and Noise Effects}
In order to observe quantum effects in our devices close to the region of the minimum operating energy gap, we need to ensure that the minimum excitation energy is much greater than $k_BT$, which for a temperature of $20\,$mK corresponds to a frequency $k_BT/h$ of about $400\,$MHz. This implies that for a thermal population of about $0.001$ at $20\,$mK, our minimum excitation energy can be approximately $3\,$GHz. This does not limit our system, as the operating frequency of interest is between $3-12\,$GHz. We also consider the repetition-time of the catapulting protocol (for $t_r\ll T_1$) to be $n T_1 = n/\Gamma_\text{tot} \approx n/\Gamma_1$ for $n \sim 5$. Here $T_1$ is the excited lifetime of the TLS. If we consider a typical value for the lifetime $T_1<200\,$ns, we can generate single microwave photons at a rate of around $10^6 \, \text{s}^{-1}$. 

Here we estimate the optimal parameters for the aforementioned CPB and flux qubit architectures by maximising the emission probability ($\eta$) and repetition-rate, while minimising the linewidth of the emitted photon ($2\pi\times\text{FWHM}=\Delta\omega = 2\Gamma_2 = \Gamma_1 + 2\Gamma_\varphi = \frac{1}{T_1} + \frac{2}{T_\phi}$). All the figures in this section are shown for an emission frequency of $6\,$GHz as it lies in the center of the band of interest. In doing these calculations we assume that the diabatic transition pulse is fast enough that decay during the pulse can be neglected. As we expect a $T_1\sim 10\,$ns while $t_r\sim 300\,$ps, we estimate that the probability of decay of the qubit during the catapult protocol to be less than $5\%$ as follows.  The total duration of the LZ protocol is $3t_r$ and the desired  photon frequency is determined by the final value $\chi_f$ of the control parameter. However, there is a duration of time ($<1.5 t_r$) during which the control parameter is not at $\chi_f$ while the qubit is excited and if the qubit were to decay at any of those $\chi \neq \chi_f$ points along the sweep, it results in a photon frequency that is not the desired frequency. This probability has an upper bound given by $1-\exp(-1.5t_r/T_1)$. For $T_1 \sim 10\,$ns and $t_r \sim 300\,$ps to be $5\%$. During these calculations we also neglect any possible non-Markovian dissipation effects that the time-dependent coupling to the environment may introduce during the LZ protocol. We also note that  since the leakage probability to the levels outside the subspace of $\ket{g}$ and $\ket{e}$ is low ($\sim 10^{-4}$, see Appendix \ref{secApp:Leakage}), the probability of a multi-photon output is low.

Beginning with the CPB, the system parameter-dependent decay \cite{burkhard_optimization_nodate} and decoherence rates\cite{koch_charge-insensitive_2007} are given by
\begin{align}
	\Gamma_{1,g/o} &= \frac{R C_{g/o}^2 \omega^2}{4C_\Sigma},\label{eqn:Gamma1_CPB}\\
	\Gamma_\phi &= \sqrt{n_\text{rms}^2 \left(\pdv{\omega}{n_g} \right)^2 + \frac{3n_\text{rms}^4}{4}\left(\pdv[2]{\omega}{n_g} \right)^2 } \nonumber\\
	&\approx \frac{n_\text{rms} E_Q}{\hbar} = \frac{n_\text{rms} (2e^2)}{\hbar C_\Sigma},\label{eqn:Gamma_phi_CPB}\\
	C_\Sigma &\approx C_g + C_o + C_J,
\end{align}
where $n_\text{rms}$ is the root mean squared (RMS) charge noise. For transmon systems, this is usually $\sim 0.5\times 10^{-3}$ \cite{zorinBackgroundChargeNoise1996}. For our system, it would be smaller considering that the charge noise is observed to decrease with decreased island size \cite{verbrugh_optimization_1995}. In addition to this, we can consider the effect of the voltage jitter of the AWG used, which corresponds to $V_\text{rms} \sim 10^{-4}\,\text{V } \equiv n_\text{rms} \sim 0.5 \times 10^{-3}$ (this approximately corresponds to a noise of $2\,\text{nV}/\sqrt{\text{Hz}}$ for a bandwidth of $3\,$GHz required for a $300\,$ps rise time). We would like to point out that $V_\text{rms}$ is mostly limited by the requirement of a large bandwidth for fast pulses, and can be decreased considerably if slower pulses are acceptable, such as in the case of small splittings at the avoided crossing.

We also observe from Eqns.~\ref{eqn:Gamma1_CPB} and \ref{eqn:Gamma_phi_CPB} for $\Gamma_1, \Gamma_\phi$  that we have a minimum possible linewidth $\Gamma_2$ if we tune $C_o$, since $\Gamma_\phi$ decreases with $C_o$ while $\Gamma_1$ increases with it. For the optimizations that we do later in this section, we take the value of $\Gamma_2$ to be close to the minimum possible value of $\Gamma_2$ achievable as we vary the output capacitance $C_o$ values, effectively allowing us to maintain low linewidths.

\begin{figure}[h]
	\centering
    \includegraphics[width=0.9\columnwidth]{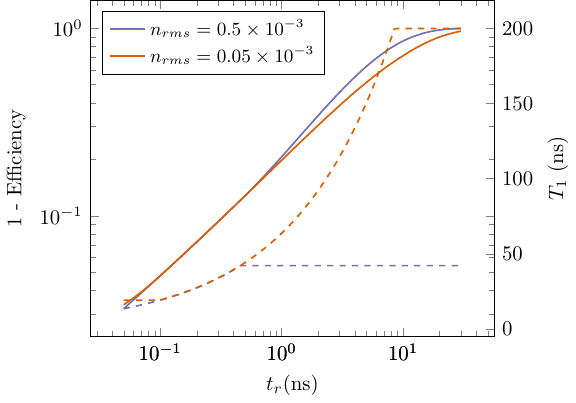}
    \caption{Variation of usable emission probability (solid) and $T_1$ (dashed) for a CPB with the rise time $t_r$ for different charge noises $n_\text{rms}$. For $n_\text{rms} = 0.5\times 10^{-3}$, linewidth is limited to $15\,$MHz while for $n_\text{rms} = 0.05\times 10^{-3}$, linewidth is limited to $5\,$MHz}
    \label{fig:CPB_opt_params}
\end{figure}

As we aim to maintain a system that can be (easily) fabricated and we want to keep the $E_Q$ as large as possible and $E_J$ as small as possible (this can be done by using a SQUID instead of a single JJ to tune the value of $E_J$ using an external magnetic field), we fix $\Gamma_\text{nr} = 1\,\mu\text{s}^{-1}$ (corresponding to a $T_{1,i} = 1\,\mu$s), $C_g = 10\,$aF, $C_J = 1\,$fF and $E_J/h = 1\,$GHz. Additionally, we would like to tune the value of $C_o$ to get the maximum usable emission probability for different sweep rates for two different charge noises, $n_\text{rms} = 0.5\times 10^{-3}$ and $n_\text{rms} = 0.05\times 10^{-3}$. As we would also like to maintain a repetition rate $\ge 1\,$MHz, we would like a $T_1 \le 200\,$ns. The optimized probabilities and $T_1$s associated with them for different charge noise rates are shown for this system in Fig.~\ref{fig:CPB_opt_params}.

We observe that for a rise time of $300\,$ps, a value that is achievable in several commercial AWGs, for a charge noise of $0.5\times 10^{-3}$ we have a usable efficiency of $90.6\,$\% with an optimal $C_o \approx 2.3\,$fF, all while maintaining a $T_1$ of $33.8\,$ns corresponding to a repetition rate of about $5.9\,$MHz (the numbers are almost the same for lower charge noise, only the linewidths are different for the emitted photon).

Next, we analyse the prospect of using flux qubits for single photon generation using our protocol. As a flux qubit's primary decoherence  channel is due to flux noise, we neglect the contribution of charge noise for this system. For our analysis, we consider the values $\Delta/h \ge 1\,$GHz and $\alpha = 0.7$ as these are shown to be achievable in fabrication \cite{deppe_superconducting_2009, yan_flux_2016}. Here, the usable efficiency is monotonically increasing with $E_J$ (or equivalently $I_c$) for a fixed $\Delta$ (See Eqns. \ref{eq:ProbFlux} and \ref{eq:EffFlux}). So, we can choose the maximum $E_J$ that is allowed by our requirements on linewidth and repetition rate. For the flux qubit, the system-dependent decay \cite{deppe_superconducting_2009, koch_charge-insensitive_2007, pengTuneableOndemandSinglephoton2016} and decoherence rates \cite{koch_charge-insensitive_2007} are approximated by,
\begin{align}
	\Gamma_{1,o} &= 2\frac{\omega Z (C_o \nu)^2}{\hbar},\\
	\Gamma_\phi &= \sqrt{\varphi_\text{rms}^2 \left(\pdv{\omega}{\phi_e} \right)^2 + \frac{3\varphi_\text{rms}^4}{4}\left(\pdv[2]{\omega}{\phi_e} \right)^2 } \nonumber\\
	&\approx \frac{\varphi_\text{rms} E_J \sqrt{1 - (1/2\alpha)^2}}{\hbar},\\
	\nu &\approx \frac{E_J}{h}\times 10^{-7}\,\textrm{V/GHz}\,
\end{align}
where the flux noise $\varphi_\text{rms} \approx 10^{-5}\Phi_0$ \cite{koch_charge-insensitive_2007}. 

\begin{figure}[h]
	\centering
    \includegraphics[width=0.9\columnwidth]{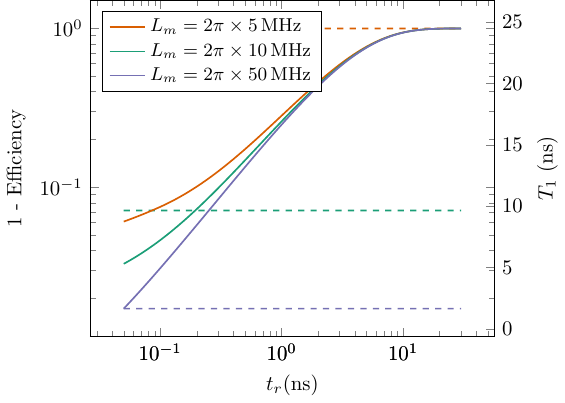}
    \caption{Variation of usable emission probability (solid) and $T_1$ (dashed) for a flux qubit with the rise time $t_r$ for different maximum allowed linewidths $L_m$}
    \label{fig:Flux_opt_params}
\end{figure}

In order to account for other sources of decay, we consider a contribution of $\Gamma_\text{nr} = 2\,\mu\text{s}^{-1}$. As we would like to have the minimum possible $\Delta$ and maximum possible $E_J$, we limit the two values to $\Delta/h = 1\,$GHz, and $E_J/h\le 250\,$GHz, and with these, once again optimize the values of $C_o$ and $E_J$ to maximise usable efficiency, while limiting the linewidth to a maximum allowed value. Finally, we also assume that, typically, a flux line can only sweep about $\pm 0.1\Phi_0$ around $0.5\Phi_0$ in a single step. With these considerations in mind, the optimized usable efficiencies and $T_1$s are shown in Fig.~\ref{fig:Flux_opt_params} for three different maximum allowed linewidths $L_m / (2\pi)$ of $50\,$MHz, $10\,$MHz and $5\,$MHz (note that for $E_J/h=250\,$GHz, the minimum linewidth is calculated to be about $2.5\,$MHz for our system).

Here, we once again see that for a rise time of $300\,$ps, we get efficiencies from about $87\textrm{--}92\,\%$ with the $T_1$ values almost being limited by the linewidths themselves, hence ranging from about $1.65\,$ns (repetition rate of $121\,$MHz) to about $25\,$ns (repetition rate of $8\,$MHz). The optimal capacitances also range from about $1.3\textrm{--}5.2\,$fF, which are achievable in fabrication.

\begin{figure}[h]
    \centering
    \includegraphics[width=0.9\linewidth]{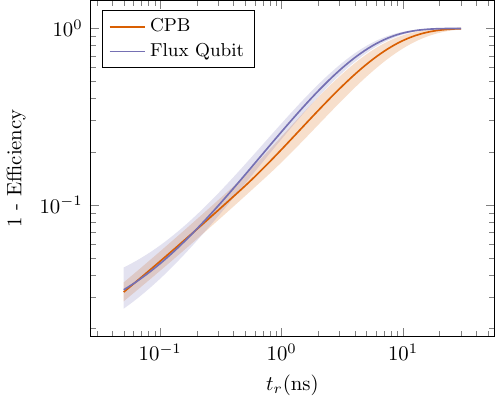}
    \caption{Effect of fabrication uncertainty on the efficiency of the CPB (orange) and the flux qubit (blue). For both, the error range consider a $10\%$ allowed uncertainty on the Josephson energy $E_J$ and the output capacitance $C_o$. The solid line denotes the ``optimal" parameters for the given limits on linewidth and repetition rate, and the shaded region of the same color represents the region of uncertainty of the ``inefficiency"}
    \label{fig:fab_uncertain}
\end{figure}

Having claimed that the parameters needed to get $\sim 90\%$ usable efficiency are feasible for physical realization through fabrication, we now examine how robust the obtained usable efficiencies are to fabrication errors. In particular, we focus on fabrication errors in the parameters $E_J$ and $C_o$ to which the device performance is most sensitive.  We allow for a $10\%$ error in $C_o$ and $E_J$ for both the flux qubit and the CPB, and plot the range as an error bar for the cases of $n_\text{rms}=0.5\times 10^{-3}$ for the CPB, and a maximum allowed linewidth of $10\ $MHz for the flux qubit in Fig.~\ref{fig:fab_uncertain}. From the figure, we can see that around a rise time of $300\ $ps, we can see that for both the CPB and the flux qubit the uncertainty in the usable efficiency is only close to $2\%$. This suggests that in addition to being able to provide reasonably high efficiencies, are reasonably robust to fabrication uncertainties.

Finally, using the obtained parameters, we can estimate the leakage of photons at the frequency of interest into the output line due to the parameter sweep. In order to account for it, consider a trapezoidal pulse of amplitude $1$ with rise time and fall time $t_r$, pulse width $T/2$, and period $T$. The envelope of the spectral function is given by ($\text{sinc}(x) = \sin(\pi x)/(\pi x)$)
\begin{equation}
    \abs{S_{trap}(\omega)} \le \frac{1}{2}\text{sinc}\left(\frac{\omega T}{4}\right) \text{sinc}\left(\frac{\omega t_r}{2}\right) \le \frac{1}{2}\text{sinc}\left(\frac{\omega T}{4}\right).
\end{equation}
Here, the average total power $\mathcal{P}$ is approximated by $(2e)^2/(C_g T)$ for a charge qubit and bounded above by $(E_J + \frac{\Phi_0^2}{2M})/T$ for the flux qubit. Using this, the power associated with the signal at $\omega \pm \Gamma$ can be approximated as $P_{l,trap} = (2\Gamma T) \mathcal{P} \abs{S_{trap}(\omega)}^2$, where $2\Gamma$ is the frequency range of interest around $\omega$ (can be taken to be the linewidth for a resonator/qubit).

Another estimate is obtained by considering a triangular pulse of amplitude $1$ with rise time and fall time $t_r$. The envelope of the spectral function is given by
\begin{equation}
    \abs{S_{tri}(\omega)} = \text{sinc}^2\left(\frac{\omega t_r}{2}\right).
\end{equation}
Once again, the power associated with the signal at $\omega \pm \Gamma$ can be approximated as $P_{l,tri} = \frac{3\Gamma t_r}{2} \mathcal{P} \abs{S_{tri}(\omega)}^2$. Here, on the other hand, $\mathcal{P}\approx (2e)^2/(C_g t_r)$ for a charge qubit, and $\approx (E_J + \frac{\Phi_0^2}{2M})/t_r$ for a flux qubit.

Using this, one can estimate that the power leakage corresponds to a photon leakage of $\beta_c P_l/(\hbar \omega \Gamma)$, where $\Gamma$ is the linewidth of our charge/flux qubit at the frequency of interest, and $\beta_c \in [0,1]$ is the factor corresponding to the coupling between the sweeping line and the output line.

With the above recommended parameters for a CPB-based system, we obtain a $P_l/(\hbar\omega\Gamma) \approx 0.07$ in the worst case of $\beta_c=1$, assuming a triangular pulse. Hence, we conclude that for a $\beta \approx \abs{2iZ\omega(C_o + C_g)}^2 \sim 10^{-5}$ for our system, this does not limit the usability of the device. Similarly, for the flux-qubit-based architecture, we obtain that the leaked photon power $P_l/(\hbar\omega\Gamma) \approx 0.97$. Despite this  comparatively large value of $P_l$ for the flux qubit, the usability of the device is not limited by the photon leakage. This is because the flux line can be decoupled even more from the output line, even with a $\beta_c \sim 10^{-3}$ (assuming that $\beta_c \approx \abs{2Z\omega C_o}^2$). 

Hence, we can see that both CPB and flux qubits are viable architectures for implementing a microwave-free single photon generation protocol with each having its own advantages. In the case where one does not have too much control over the noise in the system, we recommend the use of a flux qubit. This is mainly because the the linewidth of a flux qubit is not primarily due to $T_\phi$, but due to $T_1$ processes. On the other hand, for a CPB, $T_\phi$ determines the linewidth, thereby necessitating more complex noise mitigation strategies during design and fabrication, as well as during measurements.

\section{\label{sec:Summary}Summary and Conclusions}
In summary, we have proposed a microwave-free excitation protocol for  superconducting qubits using diabatic Landau-Zener transitions. Single microwave photons can then be generated \textit{via} spontaneous emission from the excited state.   Our theoretical estimates suggest that this protocol can be used for both Cooper-Pair Box and Flux Qubit-based architectures. Realizing robust Quantum Communication and Cryptography systems requires the high-efficiency, high-purity broadband generation of single photons. We have demonstrated that our method can be a key step in achieving this. We theoretically demonstrate a high efficiency ($>85\%$), high purity (leakage $<10^{-3}$) source with a very fast control of emitted photon frequency  over two octaves (from $\lesssim 3\,\text{GHz}$ to $\gtrsim 12\,$GHz). This fast control of photon frequencies, along with the high output rates and low footprint, makes this proposed architecture attractive for multiplexed setups where a single device can distribute its generation rate to supply photons to multiple channels separated in frequency. In addition to this, the proposed method has the advantage that it can be greatly improved upon with the development of better control electronics hardware allowing for the generation of pulses with shorter rise times, enabling even higher single-photon generation rates. While the current form of the protocol does not allow for pulse-shaping of the emitted photons, we believe that can be achieved with further studies.
\begin{acknowledgments}
The authors acknowledge the support of DST-INSPRIRE IF180339 and DST-SERB Core Research Grant CRG/2018/002129. SH acknowledges the support of the Kishore Vaigyanik Protsahan Yojana (KVPY). SH acknowledges helpful discussions with Harsh Arora and Johannes Fink.
\end{acknowledgments}
\appendix

\section{A Derivation of LZ transition Probability}
\label{secApp:LZderivation}
In this appendix, we re-derive the probability of transition for a Landau-Zener Hamiltonian as described in Ref.~\cite{zenerNonadiabaticCrossingEnergy1932}, in modern notation. The eigenstates of the system at any given point of time are $\psi_1, \psi_2$. These are also called the adiabatic basis of the system.
The initial Hamiltonian in the adiabatic basis is defined as
\begin{equation}
	H = \mqty(\sE_1 & 0\\ 0 & \sE_2).
\end{equation} 
The evolution in this basis is hard to deal with because $\psi_1, \psi_2$ evolve with $\sE_1, \sE_2$.

To make the solution easier, we consider another basis $\phi_1,\phi_2$, called the diabatic basis of the system, such that the Hamiltonian in this basis is given by
\begin{equation}
	H = \mqty(E_1 & E_{12} \\ E_{12} & E_2),
\end{equation}
and the system evolves such that 
\begin{equation}
	E_1 - E_2 = \hbar \alpha t,\ \ E_1(0) = E_0,\ \ \dot{E}_{12} = \dot{\phi_1} = \dot{\phi_2} = 0.
\end{equation} 
One also observes that at $t\to-\infty$, $\phi_1\to\psi_1$ and $\phi_2\to\psi_2$, but as $t\to\infty$, $\phi_1\to\psi_2$ and $\phi_2\to\psi_1$.

In the most general form, we can see that the Hamiltonian can be written as
\begin{equation}
	H(t) = \mqty(E_0 & E_{12} \\ E_{12} & E_0) + \mqty(\hbar \alpha/2 & 0 \\ 0 & -\hbar\alpha/2)t
\end{equation} 

So, a general state can be written as $\mqty(c_1 & c_2)^T$. Writing the Schrodinger equation, we have,
\begin{equation}\label{eq:Schrodinger}
	i\hbar \pdv{}{t}\mqty(c_1 \\ c_2) = \mqty(E_1 & E_{12} \\ E_{12} & E_2)\mqty(c_1 \\ c_2) = \mqty(E_1 c_1 + E_{12}c_2 \\ E_{12}c_1 + E_2 c_2)
\end{equation}
To convert this into a more manageable form, taking inspiration from evolution in non-perturbed systems, we write
\begin{equation}
	\mqty(c_1 \\ c_2) = \mqty(a e^{-i\int E_1 \dd t/\hbar} \\ b e^{-i\int E_2 \dd t/\hbar})
\end{equation} 
Substituting this into Eq.~\ref{eq:Schrodinger} and simplifying, we get,
\begin{equation}
    i\hbar \mqty(\dot{a} \\ \dot{b}) = E_{12} \mqty(b e^{i\int \alpha t \dd t} \\ a e^{-i\int \alpha t \dd t}) \label{eq:Der}
\end{equation}
Taking time derivatives on both sides and back-substituting into \ref{eq:Der}
\begin{align}
	0 &= \ddot{a} - i\alpha t \dot{a} + \left(\frac{E_{12}}{\hbar}\right)^2 a\\
	0 &= \ddot{b} + i\alpha t \dot{b} + \left(\frac{E_{12}}{\hbar}\right)^2 b
\end{align}
As we only care about the values at $t \to \pm \infty$ where $\ip{\phi_1}{\phi_2}=0$, we just need to solve for one of these. We will choose to solve for $b$ with the boundary conditions being
\begin{align}\label{eq:Bdry}
	\abs{c_1(-\infty)} = \abs{a(-\infty)} &= 1\\
	\abs{c_2(-\infty)} = \abs{b(-\infty)} &= 0
\end{align}
To convert the differential equation into a more manageable form (with the aim of removing the $\dot{b}$ term), we substitute $b = v e^{-(i/2)\int \alpha t \dd t}$ into \ref{eq:Bdry}. This simplifies the equation to,
\begin{equation}
	\ddot{v} + \left(\left( \frac{E_{12}}{\hbar} \right)^2 -\frac{i\alpha}{2} + \frac{\alpha^2 t^2}{4} \right)v = 0
\end{equation} 
This is a \textbf{Parabolic Cylindrical Equation}. To convert this into the standard form, we substitute
\begin{equation}\label{eq:zsubst}
	z = \sqrt{\alpha} e^{-i\pi/4}t, \ \ \xi = \frac{1}{\alpha}\left(\frac{E_{12}}{\hbar}\right)^2,
\end{equation} 
into the equation to get
\begin{equation}
	\dv[2]{v}{z} - \left( -i\xi - \frac{1}{2} + \frac{z^2}{4} \right)v = 0
\end{equation} 
Observe that $\xi$ can be used as a heuristic for the rate of the parameter sweep.

We know from Ref.~\cite{temmeAsymptoticMethodsIntegrals2015} that equations of the form:
\[
	\dv[2]{w}{z} - \left( a + \frac{z^2}{4} \right) = 0
\]
have solutions as $U(a,\pm z), V(a,\pm z), U(-a, \pm iz), V(-a, \pm iz)$. These are called \textbf{Parabolic Cylindrical Functions}.

To select which of these we want, we need to apply the second boundary condition in \ref{eq:Bdry}. As these are at $t\to -\infty$, from Eq.~\ref{eq:zsubst}, we find that $\abs{z} \to \infty$, 
\[
	\arg{z} = \begin{cases}
	            3\pi/4 & \text{if } \alpha>0\\
                -\pi/4 & \text{if } \alpha<0
	           \end{cases}
\]
In these limits, using the expansions as described in Ref.~\cite{temmeAsymptoticMethodsIntegrals2015}, we can see that $\abs{V} \ne 0$. So we only consider the functions $U(a,z)$ and its other forms.

From Ref.~\cite{temmeAsymptoticMethodsIntegrals2015},
\begin{align}
	U(a,z) \sim e^{-z^2/4}z^{-a - 1/2}\ \ &\text{if } \abs{\arg{z}}<3\pi/4\\
	U(-a, \pm iz) \sim e^{z^2/4} (\pm i)^{a - 1/2} z^{a - 1/2} \ \ &\text{if } \abs{\arg(\pm iz)}<3\pi/4
\end{align} 
As one can see, $U(a,z), U(-a, iz)$ has a singularity for $\arg{z} = 3\pi/4$, so we are forced to use $U(-a, -iz)$ as solution for $\alpha>0$. Similarly, $U(-a,-iz)$ expansion does not work for $\arg{z} = -\pi/4$, hence, we have to use either $U(a,z)$ or $U(-a,iz)$. By symmetry considerations (which is basically not wanting to solve for the prefactor twice), we use the solutions
\begin{equation}
	v(z) = N U(-a,\mp iz)\ \ \text{if }\alpha > 0, \ \alpha<0 \text{ respectively}
\end{equation}

Now, to find the value of $N$, or rather $\abs{N}$ (because to do not care about a universal phase), we make use of the first boundary condition in \ref{eq:Bdry} along with Eq.~\ref{eq:Der}. So,
\[
	\lim_{t\to -\infty} \abs{\dot{b}} = \lim_{t\to -\infty}\abs{\frac{aE_{12}}{i\hbar}} = \lim_{t\to -\infty} \abs{a} \abs{\frac{E_{12}}{\hbar}}  = \abs{\sqrt{\alpha \xi}}
\]

Also, using the definition of $v$ (solving only for $\alpha>0$ case because the other case also has the same $\abs{N}$),
\begin{align*}
	\abs{\dot{b}} &= \abs{\dot{v} - \frac{i\alpha t}{2} v} = \abs{\dv{z}{t} \dv{v}{z} - \frac{i\alpha t}{2} v}\\
	\abs{\sqrt{\xi}} &= \lim_{t\to -\infty} \abs{N} \abs{(-i) U'(-a,-iz) + \frac{z}{2} U(-a,-iz)}\\
	&= \lim_{\abs{z} \to \infty, \arg{z} = 3\pi/4} \abs{N} \abs{(-i) U'(-a,-iz) + \frac{z}{2} U(-a,-iz)}\\
	&= \abs{N} e^{\pi\xi/4}
\end{align*}
which implies,
\begin{equation}
	\abs{N} = \abs{\sqrt{\xi}} e^{-\pi\xi/4}.
\end{equation} 
Now, we just need to find the value that $\abs{b}$ takes as $t\to \infty$ which, from Eq.~\ref{eq:zsubst} correspond to $\abs{z} \to \infty$, 
\[
	\arg{z} = \begin{cases}
	-\pi/4 & \text{if }\alpha>0\\
    3\pi/4 & \text{if }\alpha<0
    \end{cases}
\]
Again, we would only be solving for the $\alpha>0$ case for the other case leads to the same form of results. For this, we use (again from Ref.~\cite{temmeAsymptoticMethodsIntegrals2015}), for $\abs{\arg{z}}\in \left( \pi/4, 5\pi/4 \right)$,
\begin{equation}
	U(a,z) \sim \frac{e^{-z^2/4}}{z^{a+1/2}} \pm \frac{i\sqrt{2\pi}}{\Gamma(a+1/2)} e^{\mp i\pi a} e^{z^2/4} z^{a-1/2}
\end{equation}

So, for using $\arg(-iz) = -3\pi/4$, we would need to use the minus sign above. Transforming, we get in the limit,
\begin{equation}
	\lim_{\abs{z} \to \infty, \arg{z} = -\pi/4} v = N \lim_{R \to \infty} \left( \frac{\sqrt{2\pi}}{\Gamma(i\xi + 1)}e^{-\pi\xi/4} e^{iR^2/4} R^{i\xi} \right)
\end{equation}
Using $\abs{\Gamma(1+ib)}^2 = \frac{\pi b}{\sinh \pi b}$, this implies,
\begin{equation}
	\lim_{\abs{z} \to \infty, \arg{z} = -\pi/4} \abs{v}^2 = 2 e^{-\pi\xi}\sinh(\pi\xi)
\end{equation} 
This is exactly the result in the Ref.~\cite{zenerNonadiabaticCrossingEnergy1932}. We can also rewrite this same result in a more revealing form to see that
\begin{equation}\label{eq:LZ}
	\lim_{t\to\infty}\abs{c_2(t)}^2 = 1 - e^{-2\pi\xi},
\end{equation}
or equivalently,
\begin{equation}
    P_\text{g $\to$ e} = e^{-2\pi\xi}
\end{equation}

\section{Introduction to the Cooper-Pair Box and Design for Photon Extraction} \label{secApp:CPBDesign}
\begin{figure}[!tb]
\centering
	\includegraphics[width=0.9\columnwidth]{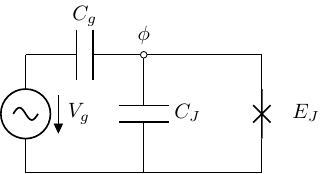}
	\caption{A gate-biased Cooper-pair box}
	\label{fig:CPB1}
\end{figure}
\begin{figure*}
\centering
	\includegraphics[width=0.9\linewidth]{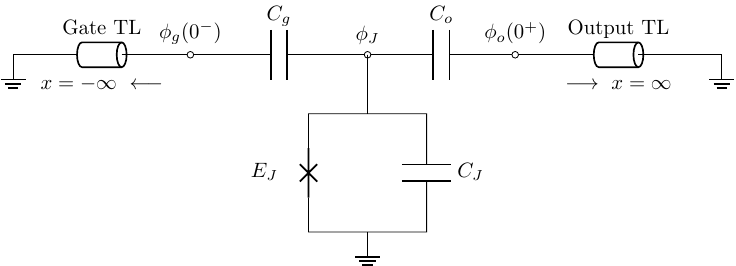}
 \caption{A gate-biased Josephson Junction with an output coupling and Input and Output lines. $C_g$ is the gate capacitance, and $C_o$ is the capacitance to the output line.}
 \label{fig:CPB2}
\end{figure*}

The classical Hamiltonian for a Cooper-pair box (Fig.~\ref{fig:CPB1}) is given by Ref.~\cite{shnirman_quantum_1997, makhlin_quantum-state_2001}
\begin{multline}
    H = \frac{(Q + C_gV_g)^2}{2(C_g + C_J)} - E_J \cos(\frac{2\pi \phi}{\Phi_0})\\*
    = E_Q \left(\frac{Q + C_g V_g}{2e}\right)^2 - E_J \cos(\frac{2\pi \phi}{\Phi_0}),
\end{multline}
where $E_Q = (2e)^2/(2C_\Sigma)$, $C_\Sigma$ is the capacitance of the island to its environment. Further defining $Q = 2ne$, $C_gV_g = -2n_g e$, and $\delta = 2\pi \phi/\Phi_0$, making use of the fact that $[\hat{Q},\hat{\phi}] = -i\hbar \equiv [\hat{\delta}, \hat{n}] = i$, we can show that up to constant terms, the quantum Hamiltonian is
\begin{multline}
    \hat{H} = E_Q\sum_{n\in\Z} (n^2 - 2n n_g)\op{n}{n} \\- \frac{E_J}{2} \sum_{n\in\Z} \left( \op{n}{n+1} + \op{n+1}{n} \right).
\end{multline}
The design of the device in Fig.~\ref{fig:CPB1}, has no way of extracting photons generated by the system. In order to allow for the extraction, we propose a device as shown in Fig.~\ref{fig:CPB2}.

Introducing the transmission lines with the extra capacitances incidentally does not change the form of the Hamiltonian (excluding the transmission line terms), but it has the effect of altering the value of $C_\Sigma$ to be given by (details derived in Appendix \ref{secApp:CPB_TLCalc}),
\begin{multline}
    C_\Sigma = C_J + \frac{C_g}{1 + \frac{k_m}{\beta_g}} + \frac{C_o}{1 + \frac{k_m}{\beta_o}}
    = C_J + C_{g,\text{eff}} + C_{o,\text{eff}},
\end{multline}
where $k_m$ is the maximum wavenumber of the signal allowed by the transmission lines (which we hypothesise would be close to the superconducting gap equivalent of the film used ($\sim 90$GHz equivalent for Aluminium, $\sim 1.1$ THz equivalent for NbTiN) as for frequencies greater than that, the dissipation is too great), $\beta_{g/o} = c/C_{g/o}$ where $c$ is the capacitance per unit length of the transmission line used (the frequency equivalent of $\beta$ is about $3.2/C$ THz where $C$ is in fF for a $50\Omega$ Coplanar Waveguide (CPW) on Silicon with $\epsilon_r\approx 11.7$).

\section{Deriving the Effect of Transmission Lines on the CPB}\label{secApp:CPB_TLCalc}
We can write the lagrangian corresponding to Fig.~\ref{fig:CPB2} as,
\begin{equation}
	\sL = \sL_{o} + \sL_{g} + \sL_{Q} + \sL_{Cg} + \sL_{Co},
\end{equation} 
where for $l,c$ being the inductance and capacitance per unit length of the transmission lines,
\begin{align}
	\sL_{o} &= \frac{1}{2}\int_{0^+}^{\infty} \dd x \left( c \dphi_o^2(x,t) - \frac{1}{l} \phi_o'^2(x,t) \right)\\
	\sL_{g} &= \frac{1}{2}\int_{-\infty}^{0^-} \dd x \left( c \dphi_g^2(x,t) - \frac{1}{l} \phi_g'^2(x,t) \right)\\
	\sL_{Q} &= \frac{1}{2}C_J \dphi_J^2 + E_J \cos(\frac{2\pi \phi_J}{\phi_0})\\
	\sL_{C_{g/o}} &= \frac{1}{2}C_{g/o} (\dphi_J - \dphi_{g/o}(0,t))^2,
\end{align}
Now, as we intend to excite the CPB using a voltage pulse on the charge line, we need to solve for the evolution of the input and output transmission line along with the CPB. In order to do so, we would first write the transmission line lagrangian in $k$-space by writing (from now on time dependence is implicitly assumed),
\begin{align}
	\phi_{g/o}(x) &= \int_0^\infty \dd k (q_{g/o}(k) \cos(kx) + p_{g/o}(k) \sin(kx)),
\end{align}
where $Z = \sqrt{l/c}$ is the characteristic impedance of the transmission lines and $q,p$ are fourier coefficients corresponding to a real $\phi$. Hence, using the orthogonality of $\cos(kx),\sin(kx)$, we can rewrite the lagrangia as (for $\omega_k = vk$, $v=1/\sqrt{lc}$, the speed of light in the transmission line),
\begin{align}
	\sL_{g/o} &= \frac{1}{2} \int_0^\infty \dd k \left( c(\ddq_{g/o}^2 + \ddp_{g/o}^2) - \frac{k^2}{l} (q_{g/o}^2 + p_{g/o}^2) \right)\\
	\sL_{C_{g/o}} &= \frac{1}{2}C_{g/o} \left(\dphi_J^2 - 2\dphi_J \int_0^\infty \dd k \ddq_{g/o}(k) \right.\nonumber\\&\quad\quad\quad\quad\quad\left.+ \left( \int_0^\infty \dd k \ddq_{g/o}(k) \right)^2 \right).
\end{align}
The primary observation of interest here is that only the $q_{g/o}$ couple with the CPB.
Now, we can write the momenta as,
\begin{align}
	\pdv{\sL}{\dphi_J} &= Q_J = C_J\dphi_J + C_o \left( \dphi_J - \int_0^\infty \dd k' \ddq_o(k') \right) \nonumber \\
	&\quad\quad\quad+ C_g \left( \dphi_J - \int_0^\infty \dd k'\ddq_g(k') \right)\\
	\pdv{\sL}{\ddq_{g/o}(k)} &= r_{g/o}(k) \nonumber\\
 &= c\ddq_{g/o}(k) - C_{g/o}  \left( \dphi_J - \int_0^\infty \dd k' \ddq_{g/o}(k')  \right)\\
	\pdv{\sL}{\ddp_{g/o}(k)} &= s_{g/o}(k) = \frac{\ddp_{g/o}(k)}{\omega_k}.
\end{align}
If one tries to solve this directly, one gets tangled up into an intractable web of infinities and zeroes. In order to alleviate us from that burden, we assume that the transmission line can only support modes up to a cutoff $k_m$ i.e. $q_{g/o}(k>k_m) = p_{g/o}(k>k_m) = 0$. 
Back-substituting $\ddq_g$ the integral and letting $\beta_J = c/C_J$, $\beta_g = c/C_g$ and $\beta_o = c/C_o$, we get,
\begin{align}
	\int_0^{k_m} \dd k' \ddq_g(k') &= \frac{1}{c (1 + \frac{k_m}{\beta_g})}\int_0^{k_m} \dd k' r_g(k') + \frac{k_m/\beta_g}{1 + \frac{k_m}{\beta_g}} \dphi_J\\
	\int_0^{k_m} \dd k' \ddq_o(k') &= \frac{1}{c (1 + \frac{k_m}{\beta_o})}\int_0^{k_m} \dd k' r_o(k') + \frac{k_m/\beta_o}{1 + \frac{k_m}{\beta_o}} \dphi_J.
\end{align}
Substituting this into the expressions for momenta, defining $I_g = \int_0^{k_m} \dd k r_g(k)$ and $I_o = \int_0^{k_m} \dd k r_o(k)$, and writing the Hamiltonian as
\begin{multline*}
    \sH = \int_0^{k_m} r_g(k) \ddq_g(k) \dd k + \int_0^{k_m} r_o(k) \ddq_o(k) \dd k \\
    + \int_0^{k_m} s_g(k) \ddp_g(k) \dd k + \int_0^{k_m} s_o(k) \ddp_o(k) \dd k + Q_J\dphi_J - \sL
\end{multline*}
and substituting and simplifying, we get,
\begin{multline}
	\sH = \frac{\left(Q_J - \left( \frac{I_g}{k_m + \beta_g} + \frac{I_o}{k_m + \beta_o} \right)\right)^2}{2c \left( \frac{1}{\beta_J} + \frac{1}{k_m + \beta_g} + \frac{1}{k_m + \beta_o} \right)} \\
	- E_J \cos(\frac{2\pi \phi_J}{\phi_0}) - \frac{I_g^2}{2c(\beta_g + k_m)} - \frac{I_o^2}{2c(\beta_o + k_m)} \\
	+ \frac{1}{2} \int_0^{k_m} \dd k \left( \frac{r_g^2(k) + s_g^2(k)}{c} + k^2\frac{q_g^2(k) + p_g^2(k)}{l}\right)\\
	+ \frac{1}{2} \int_0^{k_m} \dd k \left(\frac{ r_o^2(k) + s_o^2(k)}{c} + k^2\frac{q_o^2(k) + p_o^2(k)}{l}\right).
\end{multline}
This expression suggests that this system still acts as a charge qubit, although the coupling contributions are scaled down based on the modes allowed by the transmission line.

We can rewrite this in terms of standard quantities for a charge qubit as
\begin{multline}
	\sH = E_Q \left(N_J - N_g - N_o \right)^2 - E_J \cos(\delta) \\
	- \frac{(2e)^2N_g^2 (\beta_g + k_m)}{2c} - \frac{(2e)^2N_o^2(\beta_o + k_m)}{2c} \\
	+ \frac{1}{2} \int_0^{k_m} \dd k \left( \frac{r_g^2(k) + s_g^2(k)}{c} + k^2\frac{q_g^2(k) + p_g^2(k)}{l}\right) \\
	+ \frac{1}{2} \int_0^{k_m} \dd k \left(\frac{ r_o^2(k) + s_o^2(k)}{c} + k^2\frac{q_o^2(k) + p_o^2(k)}{l}\right).
\end{multline}
where
\begin{align}
	E_Q &= \frac{(2e)^2}{2\left( C_J + \frac{C_g}{1 + k_m/\beta_g} + \frac{C_o}{1 + k_m/\beta_o} \right)},\\
	N_{g/o} &= \frac{I_{g/o}}{k_m + \beta_{g/o}},\\
	\delta &= \frac{2\pi \phi_J}{\phi_0}.
\end{align}
Also, looking at the corresponding contributions to the charging energy, we define
\begin{align}\label{eqn:Ceff}
    C_{g/o,\text{eff}} &= \frac{C_{g/o}}{1 + k_m/\beta_{g/o}}.
\end{align}

\section{Effect of Pulse Shape on the Probability of Diabatic Excitations}\label{secApp:PulseShape}
\begin{figure}[!tb]
    \centering
    \includegraphics[width=\linewidth]{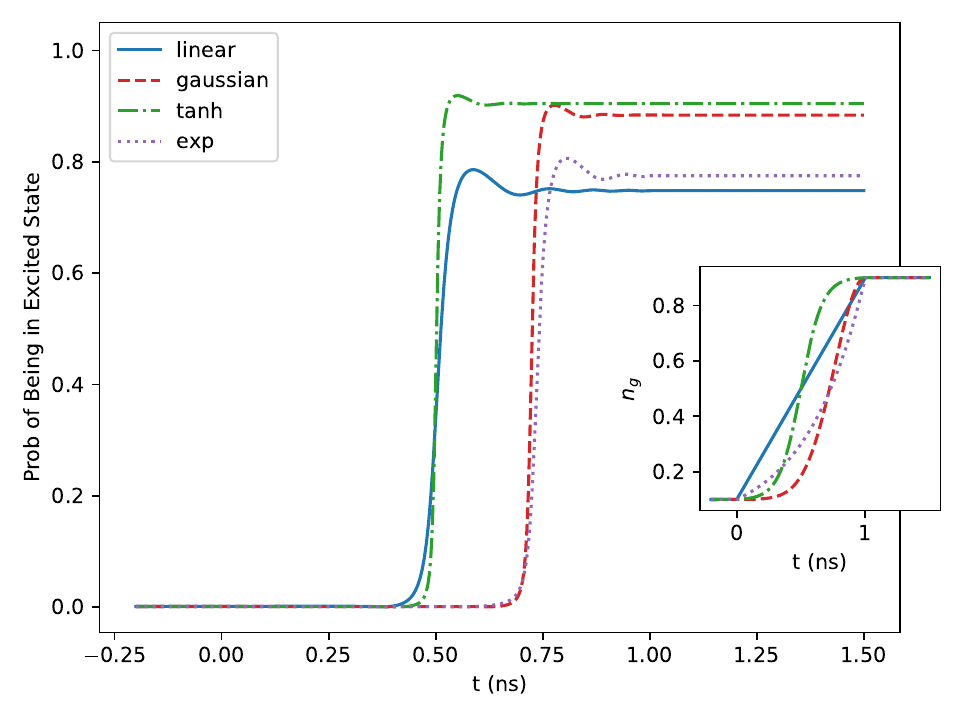}
    \caption{Simulated Evolution of the Probability of the CPB to be in the first excited state for a rise time of 1ns from $n_g=0.1$ to $n_g = 0.9$. The inset shows the pulse shapes corresponding to the rise.}
    \label{fig:PulseShape_Evol}
\end{figure}

In order to study the effect of pulse shape for sweeps that are not linear (fixed-rate), we resort to numerical simulations, as analytical expressions are difficult to obtain. For the simulations, we consider a CPB and a few common nonlinear pulse-shapes (shown in the inset of Fig.~\ref{fig:PulseShape_Evol}), namely half-gaussian ($\sigma = t_r/3$), tanh ($k = t_r/3$), and exponential, and compare them to a linear ramp-shaped pulse. All the simulations presented use $E_J/h = 1\,$GHz, $E_Q/h = 19.27\,$GHz for the CPB and consider the levels from $n=-3$ through $n=4$ for solving the Schr\"odinger equation \cite{Qutip1, Qutip2}. We note that any further increase in the number of levels does not change the simulations up to a relative error of $10^{-9}$. The pulses start at $t=0$ with $n_g=n_i$ and end at $t=t_r$ with $n_g=n_f$. For a rise time of $t_r$, the pulse shapes for $0\le t \le t_r$ are given by the expressions ($\tau = t/t_r$):
\begin{align}
    \text{Linear Rise: } & \frac{n_g(\tau) - n_i}{n_f - n_i} = \tau\\
    \text{Gaussian Rise: } & \frac{n_g(\tau) - n_i}{n_f - n_i} = \frac{e^{-9(1-\tau)^2} - e^{-9}}{1 - e^{-9}}\\
    \text{Tanh Rise: } & \frac{n_g(\tau) - n_i}{n_f - n_i} = \frac{1 + \frac{\tanh((\tau - 0.5)\times6)}{\tanh(3)}}{2}\\
    \text{Exponential Rise: } & \frac{n_g(\tau) - n_i}{n_f - n_i} = \frac{(n_f/n_i)^\tau - 1}{n_f/n_i - 1}
\end{align}

The evolution of the probability of the state not being excited for different total rise times is shown in Fig.~\ref{fig:PulseShape_Evol}. From this figure, we observe that the ``excitation" of the system happens primarily when the swept parameter $\chi\equiv n_g$ traverses the avoided crossing. This justifies the inclusion of the additional step of going from $\chi_i$ to $-\chi_0/2$ to optimize the excitation, as the probability of excitation of the system is negligible during this step.

\begin{figure}[!tb]
    \centering
    \includegraphics[width=\linewidth]{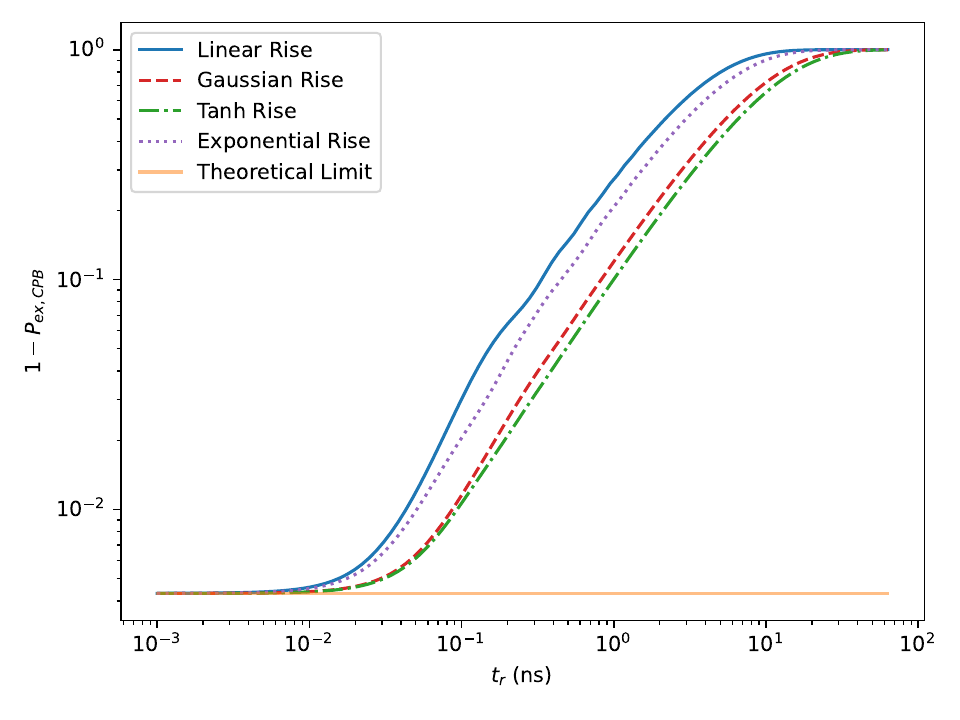}
    \caption{Simulated probability of the CPB to not be in an excited state vs rise time for different pulse shapes}
    \label{fig:PulseShape_Probs}
\end{figure}

\begin{figure}[!tb]
    \centering
    \includegraphics[width=\linewidth]{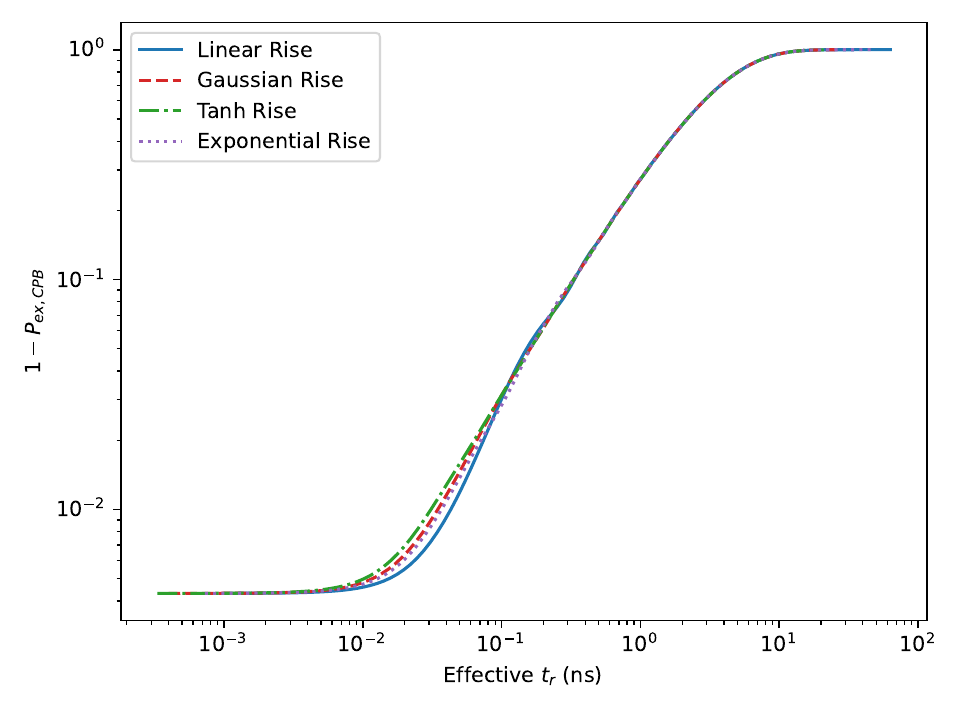}
    \caption{Simulated probability of the CPB to not be in the first excited state vs effective rise-time for different pulse shapes}
    \label{fig:PulseShape_Scaled_time}
\end{figure}

From the plot of excitation probabilities vs the rise times, as shown in Fig.~\ref{fig:PulseShape_Probs}, we observe that for the same rise time, a greater probability of excitation is achieved for the pulse shapes where the rate of sweeping is higher near the avoided crossing. This behaviour is in agreement with our expectations.
The probabilities can be well-approximated by linearising the pulses around the avoided crossing. This is clearly illustrated in the plot of excitation probabilities vs effective rise time ($t_{r,\text{eff}} = \frac{n_f - n_i}{\dot{n}_g(n_g^{-1}(0.5))}$), as shown in Fig.~\ref{fig:PulseShape_Scaled_time}. Hence, we claim that  a linear sweep is sufficient for all hardware configurations. Another important fact that we observe from the Fig.~\ref{fig:PulseShape_Probs} is that at the lowest sweep times, the probability of not being in the excited state saturates to a value of approximately $0.4\%$. This saturation can be explained by the fact that the ground state of the system at the beginning of the pulse is not exactly same as the excited state of the system at the end of the pulse for a multilevel Hamiltonian. To account for this, the "theoretical limit" line has been plotted to signify the value of $\abs{\ip{g,t=0}{e,t=t_r}}^2$, which is the maximum success probability allowed for a perfect (infinitely fast) diabatic transition.
\section{Leakage of Excitations to Outside the $\ket{g}, \ket{e}$ Subspace}\label{secApp:Leakage}
\begin{figure}[h]
    \centering
    \includegraphics[width=0.9\linewidth]{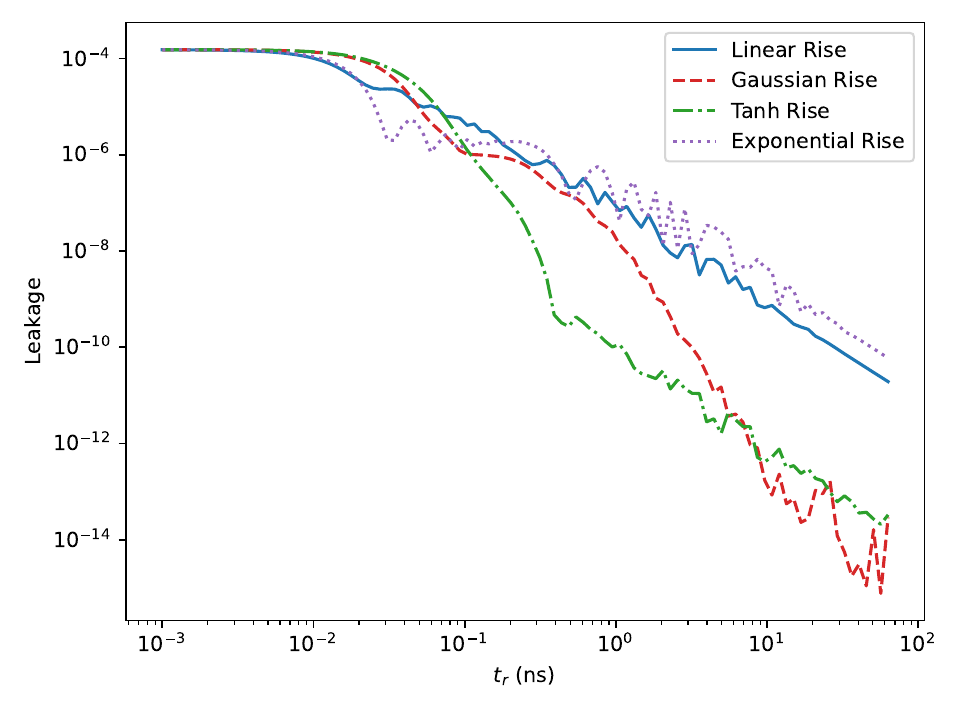}
    \caption{Effect of rise time ($t_r$) on the leakage}
    \label{fig:leakage_vs_risetime}
\end{figure}

To study the effect of the pulse sequence on leakage to outside the $\ket{g},\ket{e}$ subspace that interests us, we solve the Schr\"odinger equation for a CPB with $E_J/h = 1\,$GHz, $E_Q/h = 19.27\,$GHz considering the levels $n=-3$ through $n=4$. First, we look at the effect of rise time on the leakage for pulses that start from $n_{g,\text{min}} = 0.1$ and go to $n_{g,\text{max}} = 0.9$ (the avoided crossing occurs at $n_g = 0.5$). The results for the same are shown in Fig.~\ref{fig:leakage_vs_risetime}.

From the figure, we observe that for low enough rise-times, the leakage saturates to a low value of about $1.5\times 10^{-4}$. 
We also note that the leakage to outside the subspace contributing to multi-photon events in the output is much lower for the case of the gaussian and tanh pulses compared to linear or exponential rises. Still, we would like to point out that for currently accessible rise times of $\gtrsim 100\,$ps, the leakage to higher states due to a linear pulse is $\lesssim 10^{-6}$. 

\begin{figure}[h]
    \centering
    \includegraphics[width=0.9\linewidth]{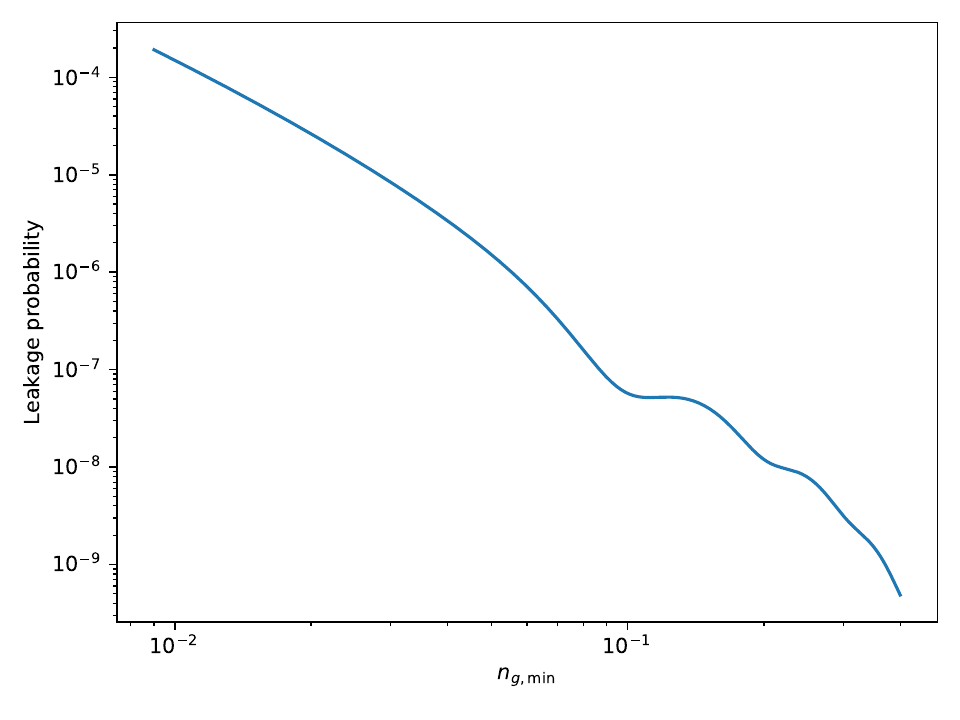}
    \caption{Effect of the range of sweep on the leakage}
    \label{fig:leakage_vs_pulserange}
\end{figure}

Having studied the effect of rise time of the pulse, we run further simulations to check the effect of the range of the sweep. We fix a rise time of $1\,$ns, and simulate the effect of pulses where $n_g$ is swept linearly from $n_{g,\text{min}}$ to $1- n_{g,\text{min}}$. The results of the simulation can be seen in Fig.~\ref{fig:leakage_vs_pulserange}. Based on these simulations we claim that it is sufficient to employ a linear pulse for a reasonably high-purity single photon generation at the output. 

\newpage

\end{document}